\documentclass[11pt]{article}
\usepackage{amsmath}
\usepackage{array}
\usepackage{tabularx}
\newcolumntype{Y}{>{\centering\arraybackslash}X}
\newcolumntype{P}[1]{>{\centering\arraybackslash}m{#1}}

\usepackage[final]{acl}

\usepackage{times}
\usepackage{latexsym}

\usepackage[T1]{fontenc}
\usepackage[utf8]{inputenc}

\usepackage{microtype}

\usepackage{amssymb}
\usepackage{inconsolata}

\usepackage{graphicx}

\usepackage{booktabs}
\usepackage{multirow}

\usepackage{xcolor}
\usepackage{colortbl}

\usepackage{tabularx}
\usepackage{makecell}

\usepackage{cuted}    % 提供 strip 环境
\usepackage{caption}  % 提供 captionof 命令

\definecolor{goodgreen}{RGB}{0, 110, 0}    
\definecolor{deltagray}{gray}{0.45}        
\definecolor{oursbg}{RGB}{242, 247, 255}   
\definecolor{gainbg}{RGB}{236, 252, 236}   

\title{AHA: Aligning Large Audio-Language Models for Reasoning Hallucinations via Counterfactual Hard Negatives}
\author{
\textbf{Yanxi Chen\textsuperscript{1*\textdagger}},
\textbf{Wenhui Zhu\textsuperscript{1*}},
\textbf{Xiwen Chen\textsuperscript{5}},
\textbf{Zhipeng Wang\textsuperscript{4}},
\textbf{Xin Li\textsuperscript{1}},
\textbf{Peijie Qiu\textsuperscript{3}},\\
\textbf{Hao Wang\textsuperscript{2}},
\textbf{Xuanzhao Dong\textsuperscript{1}},
\textbf{Yujian Xiong\textsuperscript{1}},
\textbf{Anderson Schneider\textsuperscript{5}},
\textbf{Yuriy Nevmyvaka\textsuperscript{5}},
\textbf{Yalin Wang\textsuperscript{1}}\\
\textsuperscript{1}Arizona State University,
\textsuperscript{2}Clemson University,
\textsuperscript{3}Washington University in St.Louis\\
\textsuperscript{4}Rice University,
\textsuperscript{5}Morgan Stanley\\
}

\begin{document}

\maketitle
\begin{strip}
    \centering
    \includegraphics[width=\textwidth]{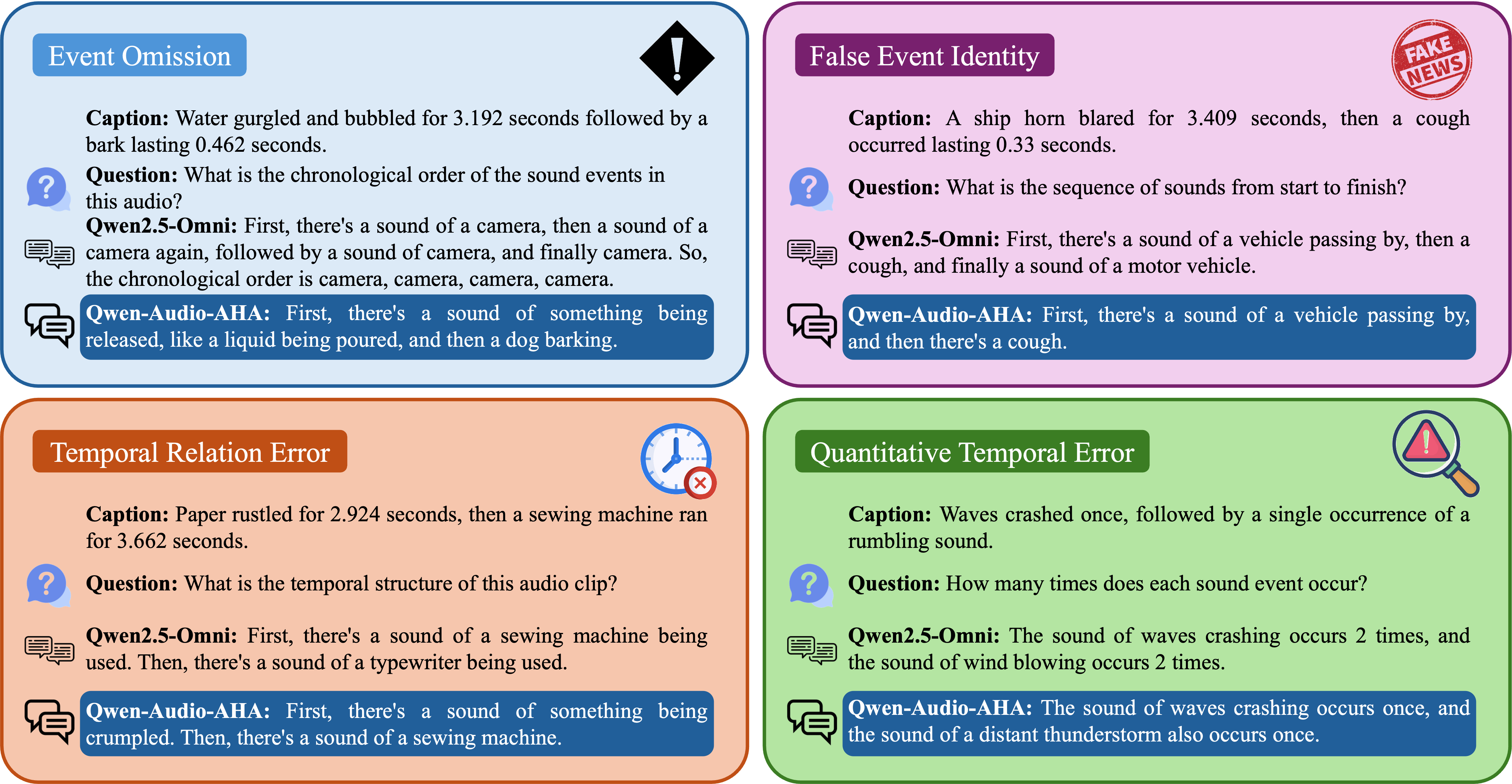}
    
    \captionof{figure}{Illustrative examples showcasing the improvements after performing the proposed \textbf{AHA framework alignment}. Compared to the base model \textbf{Qwen2.5-Omni}, our \textbf{Qwen-Audio-AHA} effectively mitigates hallucinations and errors across four critical dimensions: (1) Event Omission, (2) False Event Identity, (3) Temporal Relation Error, and (4) Quantitative Temporal Error.}
    \label{fig:case_study}
\end{strip}

\begin{abstract}

\def\thefootnote{*}\footnotetext{These authors contributed equally to this paper.}
\def\thefootnote{\textdagger}\footnotetext{Corresponding Author: ychen855@asu.edu}

Although Large Audio-Language Models (LALMs) deliver state-of-the-art (SOTA) performance, they frequently suffer from hallucinations, \textit{e.g.,} generating text not grounded in the audio input. We analyze these grounding failures and identify a distinct taxonomy: \textbf{Event Omission}, \textbf{False Event Identity}, \textbf{Temporal Relation Error}, and \textbf{Quantitative Temporal Error}. To address this, we introduce the \textbf{AHA} (\textbf{A}udio \textbf{H}allucination \textbf{A}lignment) framework. By leveraging \textbf{counterfactual hard negative mining}, our pipeline constructs a high-quality preference dataset that forces models to distinguish strict acoustic evidence from linguistically plausible fabrications. Additionally, we establish \textbf{AHA-Eval}, a diagnostic benchmark designed to rigorously test these fine-grained reasoning capabilities. We apply this data to align Qwen2.5-Omni. The resulting model, \textbf{Qwen-Audio-AHA}, achieves a 13.7\% improvement on AHA-Eval. Crucially, this benefit generalizes beyond our diagnostic set. Our model shows substantial gains on public benchmarks, including 1.3\% on MMAU-Test and 1.6\% on MMAR, outperforming latest SOTA methods. The model and dataset are open-sourced at \url{https://github.com/LLM-VLM-GSL/AHA}

\end{abstract}

\section{Introduction}
The development of Large Audio-Language Models (LALMs) has rapidly bridged the gap between acoustic perception and semantic reasoning. Foundational works like Whisper~\cite{radford2023robust} established robust speech representations, paving the way for general audio understanding models such as Qwen-Audio~\cite{chu2023qwen} and SALMONN~\cite{tang2023salmonn}, which integrate non-speech environmental cues through advanced instruction tuning. More recently, the field has shifted toward \textbf{omni-modal integration}, exemplified by GPT-4o~\cite{hurst2024gpt} and Qwen2.5-Omni~\cite{xu2025qwen2}. These models utilize simultaneous speech-text architectures to enable low-latency interaction. However, despite these architectural breakthroughs, relying solely on Supervised Fine-Tuning (SFT) leaves these models vulnerable to hallucinations~\cite{ji2023survey} and misalignment with human preferences~\cite{ouyang2022training}, often resulting in outputs that are linguistically fluent but factually detached from the acoustic reality.
Audio hallucinations present distinct challenges because sound is inherently temporal. Unlike images, acoustic events unfold over time and often overlap in complex patterns. This requires LALMs to track temporal changes with high precision. While previous studies have analyzed hallucinations, they mostly limit their scope to specific subtasks. For example, ASR research highlights silence-induced loops~\cite{koenecke2024careless,radford2023robust}, and audio captioning studies investigate fabricated objects~\cite{mei2024wavcaps}. However, the field lacks a systematic taxonomy for the complex fine-grained reasoning failures found in general-purpose LALMs. When acoustic scenes become difficult to parse, these models often default to language priors. This reliance leads to critical inaccuracies in how they ground temporal information.

To address this,
% We systematically analyze these grounding failures and define a taxonomy with four primary dimensions. 
we deconstruct these grounding failures into a distinct taxonomy with four dimensions:
\textbf{Event Omission} refers to cases where the model ignores perceptible events. \textbf{False Event Identity} occurs when the model fabricates or mislabels sounds. \textbf{Temporal Relation Error} involves distortions in chronological order or causality, such as reversing the sequence of two events. \textbf{Quantitative Temporal Error} captures incorrect event counts or duration estimates. This shift, from generic \textit{hallucination} to specific breakdowns in temporal perception and reasoning, provides the rigorous foundation necessary for our targeted alignment strategy.
% This taxonomy highlights that hallucinations in LALMs are often structural failures in \textit{temporal reasoning} rather than simple misperception. 
Crucially, the SFT struggles to fix these errors as it lacks the necessary negative constraints~\cite{ji2023survey}, which potentially encourages models to prioritize linguistic plausibility over acoustic fidelity, leading to what we term \textit{hallucinations of granularity}, where models fabricate precise details to sound coherent. We argue that effective alignment requires more than just positive examples; it demands \textit{counterfactual hard negatives}. The model must learn to distinguish the true acoustic timeline from linguistically plausible but chronologically manipulated alternatives (e.g., reversed causal events), thereby forcing it to abandon language priors in favor of strict acoustic grounding. 

To bridge this gap, we introduce the \textbf{AHA} (\textbf{A}udio \textbf{H}allucination \textbf{A}lignment) framework. This unified pipeline constructs a shared audio-question pool to generate two distinct resources. First, we synthesize a post-training alignment dataset by contrasting human-verified ground truths against ``hard negativ'' hallucinations. These negatives specifically target the error types in our taxonomy. Second, we establish AHA-Eval, a dedicated benchmark designed to rigorously assess these reasoning capabilities. Training on our alignment data forces the model to distinguish factual acoustic evidence from plausible fabrications.
Finally, we apply this framework to the Qwen2.5-Omni-7B base model by using Direct Preference Optimization (DPO), resulting in aligned model \textbf{Qwen-Audio-AHA}. Experimental results demonstrate that our framework significantly reduces error rates across all four hallucination dimensions on our diagnostic benchmark, as shown in Figure~\ref{fig:case_study}. Crucially, this improved grounding generalizes beyond the diagnostic set. \textbf{Qwen-Audio-AHA} achieves consistent accuracy gains on public benchmarks, including MMAU-Test and MMAR, demonstrating that mitigating specific hallucination behaviors contributes to broader improvements in multimodal reasoning.

% Our contributions are summarized as follows: \begin{itemize} \item We define a fine-grained taxonomy for audio hallucinations. We categorize these errors into perceptual failures and specific deficits in fine-grained reasoning. \item We introduce AHA, a unified data construction framework. It combines automated synthesis with human verification to produce a high-quality alignment dataset and a rigorous diagnostic benchmark. \item We validate the effectiveness of our data. Qwen-Audio-AHA minimizes hallucinations on diagnostic tasks and achieves superior generalization on standard public benchmarks. \end{itemize}

In summary, our contributions are fourfold: \textbf{(1)} We propose a taxonomy that divides hallucinations into four dimensions, focusing on deficits in temporal logic rather than general errors. \textbf{(2)} We introduce a diagnostic benchmark based on this taxonomy to quantify hallucinations in LALMs. \textbf{(3)} We build a preference dataset serves as a model-agnostic resource for post-training alignment and the generation pipeline also can extends to other hallucination in LALMs. \textbf{(4)} Empirical evaluations validate our framework, showing that our aligned \textbf{Qwen-Audio-AHA} significantly reduces hallucinations on diagnostic tasks with \textit{zero alignment tax}, \textbf{even boosting} performance on public benchmarks. \textbf{(5)} Our work offers a novel perspective on LALMs hallucinations, encouraging the community to shift focus to fine-grained reasoning.

\section{Related Work}

\subsection{Large Audio Language Models (LALMs)} Recent research has successfully extended Large Language Models (LLMs) into the audio domain. Whisper~\cite{radford2023robust} serves as a foundational component, providing robust representations for many subsequent systems. Early models like AudioLM~\cite{borsos2023audiolm}, SpeechGPT~\cite{zhang2023speechgpt}, and AudioPaLM~\cite{rubenstein2023audiopalm} treated audio as discrete tokens or utilized joint vocabularies for unified sequence modeling. Later work expanded this scope to general audio understanding. For instance, SALMONN~\cite{tang2023salmonn} combined encoders to capture environmental cues, while Qwen-Audio~\cite{chu2023qwen} scaled up analysis through multi-task instruction tuning. Additionally, WavLLM~\cite{hu2024wavllm} introduced dual-encoder architectures to improve context robustness.
Current trends prioritize \textbf{omni-modal integration}, where models learn directly from mixed data streams~\cite{hurst2024gpt,defossez2024moshi,xie2024mini}. Qwen2.5-Omni~\cite{xu2025qwen2} exemplifies this shift, utilizing a ``Thinker-Talker'' architecture for low-latency interaction. However, relying solely on Supervised Fine-Tuning (SFT) leaves these models prone to hallucinations~\cite{ji2023survey} and misalignment with human preferences~\cite{ouyang2022training}. To address this, we introduce AHA, a unified framework designed for post-training alignment that seamlessly integrates with existing LALMs. It mitigates temporal hallucinations through a targeted training dataset and provides a diagnostic benchmark to rigorously evaluate fine-grained reasoning capabilities.
\begin{figure*}[t]
    \centering
    \includegraphics[width=\linewidth]{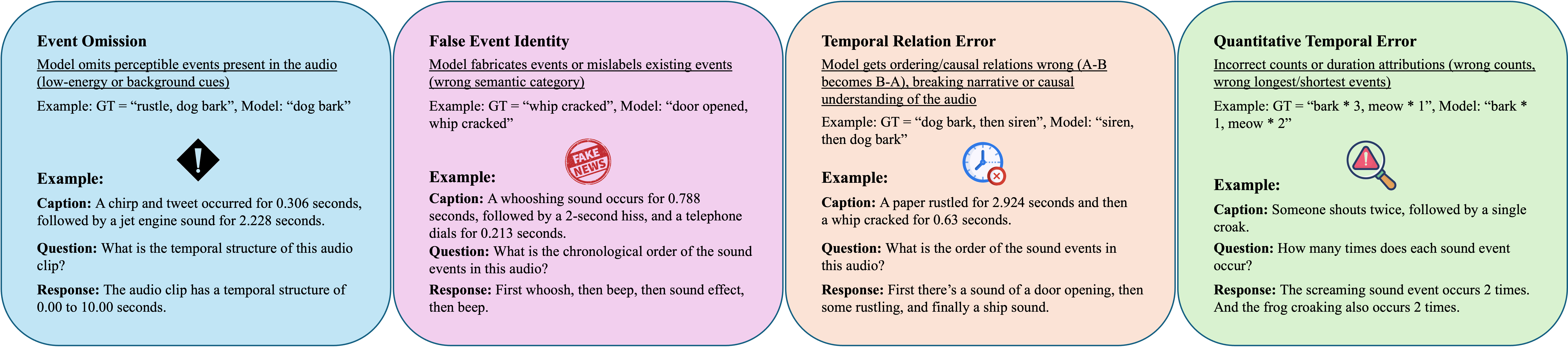}
    \caption{Four principal hallucination types observed in temporal audio reasoning: Event Omission, False Event Identity, Temporal Relation Error, and Quantitative Temporal Error. Each panel illustrates the definition of the failure mode along with a representative hallucination example produced by Qwen2.5-Omni on diagnostic testing.}
    \label{fig:motivation}
\end{figure*}

\subsection{Hallucination in LALMs}
LALMs have made strides but remain prone to hallucinations ungrounded in acoustic signals. This issue is uniquely challenging in the audio domain because sound events are temporally dynamic and often overlapping. Previous research has primarily addressed isolated subtasks. For instance, ASR studies characterize silence-induced loops~\cite{radford2023robust,koenecke2024careless,guerreiro2023hallucinations,frieske2024hallucinations}, while audio captioning work investigates object fabrication~\cite{mei2024wavcaps,gong2023listen}. However, the field lacks a systematic taxonomy for the complex fine-grained reasoning failures in general LALMs. To fill this gap, we categorize these errors into four dimensions: \textbf{Event Omission}, \textbf{False Event Identity}, \textbf{Temporal Relation Error}, and \textbf{Quantitative Temporal Error}. Standard SFT often fails to rectify such granular errors because it lacks negative constraints~\cite{ji2023survey}. We address this by introducing a specialized dataset and benchmark designed for effective post-training alignment.

\section{Problem Statement}
Formally, we define an LALM as a function that maps an acoustic sequence $\mathcal{A}$ and a textual instruction $\mathcal{I}$ to a response $\mathcal{R}$. Temporal audio reasoning, however, poses challenges that extend far beyond transcription or high-level acoustic captioning. A model must accurately determine what events occur, when they occur (temporal order), how long they last (duration), and how often they appear (count). Crucially, any generated response $\mathcal{R}$ should remain strictly grounded in the acoustic evidence provided by $\mathcal{A}$. Our diagnostic analysis of state-of-the-art omni-modal LALMs (e.g., Qwen2.5-Omni) reveals that current systems frequently violate this grounding requirement. Instead of relying on the actual audio signal, the model often falls back on language priors, generating responses that are semantically plausible yet acoustically incorrect. These deviations manifest through a compact set of systematic failure modes that directly affect temporal grounding and degrade performance on downstream applications such as acoustic monitoring, multimedia forensics, and event analytics. To characterize these behaviors, we introduce a four-category taxonomy of hallucinations that captures the dominant and practically actionable errors exhibited by contemporary LALMs. This taxonomy aligns closely with the controlled perturbations introduced by our data generation pipeline, enabling precise attribution and measurement. Notably, even the SOTA LALM we evaluated demonstrates all categories of these failure modes.

\begin{figure*}[t]
    \centering
    \includegraphics[width=0.9\linewidth]{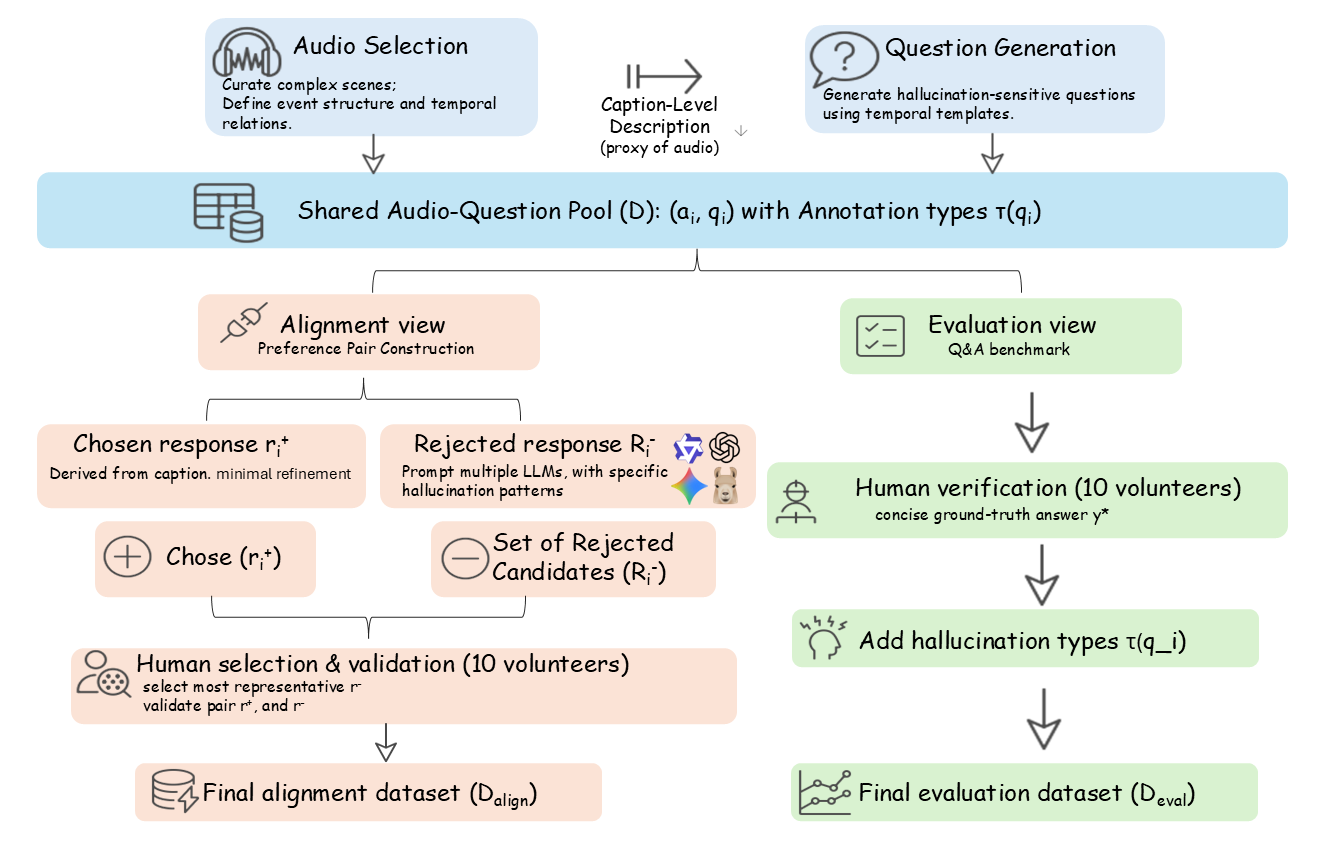}
    \caption{\textbf{The unified data construction pipeline for AHA.} The process begins by establishing a shared \textbf{Audio-Question Pool ($\mathcal{D}$)} derived from complex acoustic scenes and hallucination-sensitive fine-grained reasoning templates. From this foundation, the pipeline bifurcates into two complementary views: (Left) The \textbf{Alignment View} constructs preference pairs for DPO by contrasting caption-derived chosen responses ($r_i^+$) against LLM-generated rejected responses ($r_i^-$) that contain specific hallucination patterns. (Right) The \textbf{Evaluation View} establishes a rigorous QA benchmark by collecting human-verified ground-truth ($y_i^*$) and annotating fine-grained hallucination types ($\tau(q_i)$).}
    \label{fig:frame}
    % \vspace{-0.5cm}
\end{figure*}

\noindent \textcolor{blue}{\textit{\textbf{-Event Omission:}}} The model fails to mention perceptible events present in the audio (e.g., a faint hiss before a loud impact is ignored). Omission harms recall of the acoustic scene and often reflects attention bias toward high-energy events. \\
\noindent \textcolor{blue}{\textit{\textbf{-False Event Identity (Fabrication \& Misclassification):}}} The model either (a) fabricates events that are not present in the audio or (b) assigns the wrong semantic label to an existing event (e.g., calling a cap gun a “whip” or inventing a “door opening” that never occurs). Both behaviors reflect errors in mapping acoustic evidence to semantic labels. \\
\noindent \textcolor{blue}{\textit{\textbf{-Temporal Relation Error:}}} The model gets relations between events wrong: it swaps the chronological order, or misstates causal/sequence relations (e.g., saying B$\rightarrow$A when A$\rightarrow$B). These errors break narrative or causal understanding of the audio. \\
\noindent \textcolor{blue}{\textit{\textbf{-Quantitative Temporal Error:}}} The model errs on quantitative temporal attributes: incorrect counts of repeated events (count bias) or wrong comparative duration attributions (e.g., saying the shorter event is longest). These mistakes indicate failure in simple temporal arithmetic and duration comparison

We quantify these deficiencies using a specialized diagnostic evaluation. Figure~\ref{fig:motivation} reveals that Qwen2.5-Omni struggles significantly across all dimensions, showing an Omission Rate over 40\% and a Boundary IoU under 0.4. These results confirm that standard SFT is inadequate for complex acoustic scenes. The model frequently defaults to language priors rather than maintaining strict auditory grounding. This structural weakness necessitates a post-training strategy that penalizes ungrounded generation. Motivated by this, we propose AHA, a framework designed to enforce alignment and mitigate these complex fine-grained reasoning failures.

\section{AHA: A Unified Post-Training Framework for Audio Hallucination}
\label{sec:AHA}

\subsection{Dataset Formulation and Unified Design}
\label{subsec:aha-formulation}

We introduce \textbf{AHA}, a unified dataset designed to both \emph{align} and \emph{evaluate} LALMs on hallucination behaviors arising from complex acoustic scenes.  
While AHA is grounded in audio data, a key design consideration is that many existing LLMs do not directly accept audio as input. Instead, these models typically operate on textual descriptions or instructions derived from audio content.
Accordingly, AHA is constructed around caption-level representations that serve as a proxy for acoustic input, enabling the study and mitigation of audio-related hallucinations in a model-agnostic manner.
All audio samples in AHA are sourced from existing captioned audio corpora, specifically \textbf{\textit{AudioTime}}~\cite{xie2025audiotime}, which contains 20,000 training and 2,000 testing data samples in four categories: duration, frequency, ordering and timestamp. Each data sample consists of a temporally structured audio clip paired with a descriptive caption. We used a subset of AudioTime as our data source, which contains all samples with multiple audio events in the audio clip, resulting in 11,507 training and 1,251 testing samples.
Formally, AHA is defined over a shared set of audio-question pairs:
\begin{equation}
\mathcal{D} = \{(a_i, q_i)\}_{i=1}^{N},
\end{equation}
where $a_i$ denotes an audio clip and $q_i$ denotes a hallucination-sensitive question generated from its caption-level description.  
The audio signal $a_i$ is used to establish ground truth and human verification, whereas model-facing supervision is provided through caption-derived instructions.
From this shared pool, we derive two complementary views:
\begin{align}
\mathcal{D}^{\text{align}} &= \{(a_i, q_i, r_i^+, r_i^-)\}, \\
\mathcal{D}^{\text{eval}} &= \{(a_i, q_i, y_i^*, \tau(q_i))\},
\end{align}
where $r_i^+$ and $r_i^-$ denote preferred and dispreferred responses, $y_i^*$ is a ground-truth answer, and $\tau(q_i)$ denotes hallucination type annotations.  
Both views share the same underlying audio samples, questions, and hallucination taxonomy, ensuring that alignment and evaluation target consistent hallucination phenomena.

\subsection{Hallucination-Oriented Construction}
\label{subsec:aha-construction}

\paragraph{Audio selection.}
Audio samples are curated from existing captioned audio corpora with an explicit focus on hallucination-prone scenarios. We prioritize clips that contain multiple sequential or overlapping sound events, frequent scene transitions, and non-trivial temporal dependencies.  
Although the audio signal is not directly consumed by most LLMs, it serves as the authoritative source for defining event structure, temporal relations, and annotation validity.

\paragraph{Question generation.}
Given a caption-level description of an audio clip, we programmatically generate hallucination-sensitive questions by sampling from a predefined set of \emph{fine-grained reasoning templates}.  
These templates are designed to elicit hallucination behaviors commonly observed when models reason over incomplete or abstracted representations of audio content.
The templates are organized into the following categories:
\textbf{1) Explicit temporal ordering}, e.g., identifying first/last events or exact event sequences;
\textbf{2) Temporal logic and counterfactual reasoning}, e.g., order verification or hypothetical trimming;
\textbf{3) Temporal counting and duration comparison}, e.g., event frequency or longest/shortest event.
Each template may include symbolic placeholders (e.g., \emph{Event A}, \emph{Event B}), which are instantiated using event mentions inferred from the caption.  
Formally, each question $q_i$ is associated with one or more hallucination types:
\begin{equation}
\begin{aligned}
\tau(q_i) \subseteq \Big\{&
\textsc{Omission},\
\textsc{FalseIdentity},\\
&\textsc{TemporalOrder},\
\textsc{Quantitative}
\Big\}.
\end{aligned}
\end{equation}

These questions serve as a shared intermediate representation reused by both the alignment and evaluation views of AHA, enabling consistent supervision despite differences in model input modalities. (\textcolor{red}{see Appendix~\ref{appendix:2.3} for the list of question templates})

\subsection{Alignment: Counterfactual Hard Negative Synthesis}
\label{subsec:aha-align}

We construct preference pairs via counterfactual hard negative synthesis.
For each audio-question pair $(a_i, q_i)$, the \textbf{chosen response} $r_i^+$ derives directly from the ground truth caption to ensure factual accuracy.
To generate the \textbf{rejected responses}, we employ a structured prompting strategy using external text-only LLMs.
We explicitly instruct the model to synthesize counterfactual errors corresponding to the hallucination taxonomy defined in Section~\ref{subsec:aha-construction}.
For instance, the prompt directs the model to swap the chronological order of events, modify specific counts, or fabricate non-existent sounds based on the ground truth (\textcolor{red}{see Appendix~\ref{appendix:2.3} for prompt templates}).
This process yields a set of negative candidates:
\begin{equation}
\mathcal{R}_i^- = \{r_i^{-(1)}, \dots, r_i^{-(K)}\}.
\end{equation}
These candidates serve as \textbf{hard negatives} because they mimic the linguistic fluency of the chosen response while strictly contradicting the temporal logic of the audio.

\paragraph{Human selection and validation.}
From the candidate set $\mathcal{R}_i^-$, \textbf{10 human volunteers} with experience in audio understanding select the single most representative hallucinated response, denoted as $r_i^-$.
Selection criteria prioritize responses that are linguistically plausible under caption-only reasoning yet clearly incorrect given the acoustic evidence.
This ensures the final pair provides a sharp contrast for optimization.
The resulting alignment dataset is formalized as:
\begin{equation}
\mathcal{D}^{\text{align}} = \{(a_i, q_i, r_i^+, r_i^-)\}_{i=1}^{N_{\text{align}}}.
\end{equation}
This preference-based supervision provides a reliable signal for Direct Preference Optimization (DPO), explicitly teaching models to prefer fine-grained reasoning over hallucinations.

\subsection{AHA Eval: Hallucination QA Benchmark}
\label{subsec:aha-eval}

The evaluation view of AHA reuses the same audio-question pairs as the alignment view but differs in supervision format.  
For a subset of $(a_i, q_i)$, a group of \textbf{10 human volunteers} manually verify a concise ground-truth answer:
\begin{equation}
y_i^* = \mathrm{Ans}(a_i, q_i),
\end{equation}
which is strictly supported by perceptible acoustic evidence and avoids unnecessary temporal metadata unless explicitly required by the question.
Each evaluation instance is additionally annotated with hallucination types $\tau(q_i)$, enabling fine-grained, category-level diagnostic evaluation:
\begin{equation}
\mathcal{D}^{\text{eval}} = \{(a_i, q_i, y_i^*, \tau(q_i))\}_{i=1}^{N_{\text{eval}}}.
\end{equation}

\section{Experiments \& Results}
We evaluate our alignment approach across a mixture of in-domain and out-of-domain benchmarks to assess both (A) reduction in hallucinations and (B) general improvements in fine-grained reasoning and broader audio understanding. We compare 6 models: Qwen2.5-Omni (base), Qwen2.5-Omni-DPO ($\beta$=0.3), Kimi-Audio, Audio Flamingo 3, GPT-4o-mini and Gemini-2.5. Experiments use both AHA-Eval and a suite of public audio benchmarks: MMAU-test, MMAU-test-mini, MMAU-Pro, and MMAR.

\subsection{Alignment Post-Training}
\label{subsec:alignment-training}

We perform preference alignment on \textbf{Qwen2.5-Omni} using our dataset $\mathcal{D}^{\text{align}}$. We adopt DPO~\cite{rafailov2023direct} to steer the model toward audio-grounded reasoning. Unlike RLHF, DPO optimizes the policy directly on preference data without requiring a separate reward model.
Let $x = (a, q)$ denote the audio-question input. For each instance, the dataset provides a preferred response $r^+$ (ground-truth) and a rejected response $r^-$ (hallucination). We optimize the policy $\pi_{\theta}$ against the frozen reference model $\pi_{\text{ref}}$ by minimizing:

\begin{equation}
\label{eq:dpo_loss}
\resizebox{1.0\linewidth}{!}{$
\begin{aligned}
\mathcal{L}_{\text{DPO}}(\pi_{\theta}; \pi_{\text{ref}}) = -\mathbb{E}_{(x, r^+, r^-) \sim \mathcal{D}^{\text{align}}} \Big[ \log \sigma \Big( & \beta \log \frac{\pi_{\theta}(r^+ | x)}{\pi_{\text{ref}}(r^+ | x)} \\
& - \beta \log \frac{\pi_{\theta}(r^- | x)}{\pi_{\text{ref}}(r^- | x)} \Big) \Big]
\end{aligned}
$}
\end{equation}
where $\sigma$ is the sigmoid function and $\beta$ controls the divergence from $\pi_{\text{ref}}$. This objective increases the likelihood of grounded responses while explicitly suppressing the specific hallucination in $r^-$.

\begin{table*}[t]
    \centering
    \caption{\textbf{Unified Evaluation Results.} 
    \textbf{Panel A} reports hallucination error rates (\%) on the AHA benchmark (\textbf{\underline{lower is better}}).
    \textbf{Panel B} reports accuracy (\%) on public benchmarks (\textbf{\underline{higher is better}}). 
    \colorbox{oursbg}{Blue rows} highlight our aligned model's absolute performance.
    \colorbox{gainbg}{Green rows} quantify the substantial improvement ($\Delta$) over the baseline.
    }
    \label{tab:unified_results}
    
    % 使用 tabularx 填满文本宽度，不再需要 resizebox (字体更清晰)
    % 如果觉得表格太长，可以在外面套 resizebox，但通常不建议缩放 tabularx
    \small % 稍微减小字号以适应内容
    \begin{tabularx}{\textwidth}{l Y Y Y Y}
        \toprule
        % ================= PANEL A =================
        \multicolumn{5}{c}{\textbf{\textsc{Panel A: AHA Eval Benchmark (Error Rate \%, Lower is Better)}}} \\
        \cmidrule(lr){1-5}
        % 使用 \makecell 实现表头换行，大大缩减列宽
        \textbf{Model} & \textbf{\makecell{Event\\Omission}} & \textbf{\makecell{False\\Identity}} & \textbf{\makecell{Temp.\\Relation}} & \textbf{\makecell{Quant.\\Temporal}} \\
        \midrule
        GPT-4o & 77.8 & 72.2 & 34.1 & 63.6 \\
        Gemini 2.5 & 74.7 & 80.8 & 34.8 & \underline{58.3} \\
        Kimi-Audio & 84.7 & \textbf{57.6} & 35.6 & 72.5 \\
        Audio Flamingo 3 & \underline{61.4} & 81.6 & 40.0 & 75.3 \\
        
        \midrule 
        
        Qwen2.5-Omni (Base) & 70.6 & 70.6 & \underline{30.5} & 69.6 \\
        
        % === Ours Model Row ===
        \rowcolor{oursbg}
        \textbf{Qwen-Audio-AHA (Ours)} & \textbf{53.8} & \underline{64.1} & \textbf{15.9} & \textbf{52.6} \\
        
        % === Delta Row ===
        \rowcolor{gainbg}
        \multicolumn{1}{r}{\textit{\footnotesize Reduction ($\Delta$)}} & 
        \footnotesize \textcolor{goodgreen}{$\blacktriangledown$ \textbf{-16.8}} & 
        \footnotesize \textcolor{goodgreen}{$\blacktriangledown$ \textbf{-6.5}} & 
        \footnotesize \textcolor{goodgreen}{$\blacktriangledown$ \textbf{-14.6}} & 
        \footnotesize \textcolor{goodgreen}{$\blacktriangledown$ \textbf{-17.0}} \\
        
        \midrule
        \midrule
        
        % ================= PANEL B =================
        \multicolumn{5}{c}{\textbf{\textsc{Panel B: Public Benchmarks (Accuracy \%, Higher is Better)}}} \\
        \cmidrule(lr){1-5}
        % 同样对长标题进行换行处理，保持整齐
        \textbf{Model} & \textbf{\makecell{MMAU-test-mini}} & \textbf{\makecell{MMAU-test}} & \textbf{\makecell{MMAU-Pro}} & \textbf{MMAR} \\
        \midrule
        GPT-4o & 57.8 & 54.5 & 41.7 & 49.2 \\
        Gemini 2.5 & 70.5 & 67.1 & \textbf{54.9} & 57.3 \\
        Kimi-Audio & 54.9 & 56.1 & 50.1 & 49.1 \\
        Audio Flamingo 3 & 74.5 & \underline{72.4} & 25.0 & 54.1 \\
        
        \midrule 
        
        Qwen2.5-Omni (Base) & \underline{74.6} & 71.2 & 54.2 & \underline{58.1} \\
        
        % === Ours Model Row ===
        \rowcolor{oursbg}
        \textbf{Qwen-Audio-AHA (Ours)} & \textbf{76.4} & \textbf{72.5} & \underline{54.8} & \textbf{59.7} \\
        
        % === Delta Row ===
        \rowcolor{gainbg}
        \multicolumn{1}{r}{\textit{\footnotesize Improvement ($\Delta$)}} & 
        \footnotesize \textcolor{goodgreen}{$\blacktriangle$ \textbf{+1.8}} & 
        \footnotesize \textcolor{goodgreen}{$\blacktriangle$ \textbf{+1.3}} & 
        \footnotesize \textcolor{goodgreen}{$\blacktriangle$ \textbf{+0.6}} & 
        \footnotesize \textcolor{goodgreen}{$\blacktriangle$ \textbf{+1.6}} \\
        
        \bottomrule
    \end{tabularx}
\end{table*}

\paragraph{Implementation Details.}
We use Qwen2.5-Omni-7B as our foundation model. To reduce computational cost, we apply Low-rank adaptation (LoRA)~\cite{hu2022lora} in our DPO post-training. Following the effective hyperparameter choices for LoRA~\cite{schulman2025lora}, We set the rank $r = \text{16}$ and scaling factor $\alpha = \text{32}$, and attached LoRA layers to all attention and MLP layer in the model. The reference model is frozen in 16-bit precision. We train for $\text{8}$ epochs using the AdamW optimizer with a learning rate of $2\times10^{-6}$ and a global batch size of $16$. The DPO parameter $\beta$ is set to $\text{0.3}$. Experiments are conducted on $\text{4}$ NVIDIA A100 GPUs.
For brevity, we refer to this aligned model as \textbf{Qwen-Audio-AHA} in the subsequent sections. \textcolor{red}{(More details refer to appendix~\ref{appendix:1})}

\subsection{AHA Benchmark Evaluation}
AHA provides a controlled setting to quantify hallucination behaviors arising from complex fine-grained reasoning (Section 4). Each test instance is annotated with one or more of the taxonomy-defined hallucination types. We leverage LLM-as-a-judge  by providing ground-truth audio captions to a GPT-4o client, which then label model responses across the four dimensions. This allows us to calculate precise metrics: \textbf{event omission rate}, \textbf{false-identity rate}, \textbf{ordering error rate} and \textbf{counting error rate}. We report hallucination rates of the generated questions, which follow the unified design of AHA (Sections 4.1–4.4).
As shown in Table~\ref{tab:unified_results} (Panel A), current SOTA models exhibit severe grounding failures. Even strong baselines like Gemini and GPT-4o exceed 74\% in Event Omission rates. Kimi-Audio, while robust in identity, suffers an 84.7\% omission rate. These results underscore that standard training paradigms often fail to ensure strict adherence to acoustic evidence.

\paragraph{Hallucination Reduction.} 
Across all four hallucination dimensions, Qwen-Audio-AHA substantially reduces error rates compared to the Qwen2.5-Omni base model. The most significant absolute gains were observed in Quantitative Temporal Error and Event Omission, where we achieved reductions of 17.0\% and 16.8\% respectively.
In quantitative fine-grained reasoning, where baselines often return incorrect event frequencies or durations, our aligned model lowered the error rate from 69.6\% to 52.6\%. Similarly, in Temporal Relation reasoning, the most challenging category for many models, Qwen-Audio-AHA nearly halved the base model's error rate, dropping it from 30.5\% to 15.9\%. These results indicate that DPO effectively teaches the model to prioritize acoustic grounding over the linguistically plausible but factually incorrect priors found in caption-only models (Table~\ref{tab:unified_results}).

\paragraph{Quality of Fine-Grained Reasoning.} To illustrate the impact of alignment on model output, we conducted qualitative case studies using the AHA dataset. We found that Qwen-Audio-AHA consistently produces more concise and acoustically faithful responses. While the base model frequently exhibits ``hallucinations of granularity'', such as fabricating overlapping events or misstating causal sequences, the aligned model generates results that adherence to the temporal metadata derived from the audio. This finding motivates using AHA as a diagnostic hallucination benchmark, where precise event tuples extracted from captions allow scoring.

\subsection{Generalization to Public Benchmarks}
To evaluate the impact of our alignment on broad multimodal audio understanding, we evaluated \textbf{Qwen-Audio-AHA} on four standard benchmarks: MMAU-test, MMAU-test-mini, MMAU-Pro, and MMAR. We strictly adhered to official protocols for all datasets.

\paragraph{Performance Consistency and Mutual Benefits.}
Table~\ref{tab:unified_results} (Panel B) presents the evaluation results. It is important to note a \textit{granularity mismatch}: while AHA targets \textit{fine-grained} temporal hallucinations, public benchmarks primarily assess \textit{coarse-grained} perception, which may dilute the visibility of our improvements. Typically, such targeted optimization incurs an ``alignment tax''~\citep{ouyang2022training}, where general capabilities degrade under specific constraints. However, \textbf{Qwen-Audio-AHA} defies this trend, exhibiting \textit{positive transfer} across the board. We observe consistent accuracy gains: \textbf{+1.8\%} on MMAU-test-mini, \textbf{+1.3\%} on MMAU-test, and \textbf{+1.0\%} on MMAR. These metrics confirm that our framework enhances model precision on general tasks without overfitting to the preference dataset.

\paragraph{Generalization of Fine-grained Reasoning Capabilities.}
Notably, the reasoning capabilities learned from our alignment transfer effectively to public benchmarks. We observe significant improvements in tasks that require complex reasoning. This indicates that the \textbf{AHA} framework does not simply memorize caption templates. Instead, it equips the model with robust \textbf{complex fine-grained reasoning}. As a result, the model achieves better performance on diverse, out-of-distribution public benchmarks.
% \begin{figure*}[t]
%     \centering
%     \includegraphics[width=\linewidth]{analysis.png}
%     \caption{Case study of the four dimensions before and after DPO alignment: 1) Event Omission, 2) False Event Identity, 3) Temporal Relation Error, 4) Quantitative Temporal Error. At evaluation time, we leveraged LLM-as-a-judge method by providing the ground-truth captions to the judge model.}
%     \label{fig:case_study}
% \end{figure*}

\section{Analysis}
\paragraph{\textit{Does hallucination alignment improve general reasoning?}} Contrary to concerns that alignment might degrade general capabilities, our results indicate the opposite. As detailed in Table~\ref{tab:temporal_results}, Qwen-Audio-AHA achieves substantial gains on the \textbf{Temporal Event Reasoning} subsets of public benchmarks, peaking at \textbf{+8.3\%} on MMAU-Test. Notably, these specific gains far outpace the model's average improvement on general benchmarks (+0.6\% to +1.8\%). This discrepancy confirms that our method does not merely suppress text, but it also refines the internal representation of the audio timeline. Consequently, the model becomes far more effective at resolving complex event sequences that previously triggered ``hallucinations of granularity''.

\begin{table}[htp]
    \centering
    
    \caption{\textbf{Complex Temporal/Event Reasoning Accuracies.}
    \colorbox{oursbg}{Blue}: Ours.
    \colorbox{gainbg}{Green}: Improvement ($\Delta$).
    }
    \label{tab:temporal_results}

    \resizebox{\columnwidth}{!}{
        \begin{tabular}{l c c c}
            \toprule
            
            \multicolumn{4}{c}{\textbf{\textsc{Complex Fine-Grained Reasoning Accuracies (Higher is Better)}}} \\
            \cmidrule(lr){1-4}

            \textbf{Model} & \textbf{MMAU-Mini} & \textbf{MMAU-Test} & \textbf{MMAU-Pro} \\
            \midrule
            
            Qwen2.5-Omni (Base) & \underline{70.8} & \underline{59.5} & \underline{64.4} \\

            \rowcolor{oursbg}
            \textbf{Qwen-Audio-AHA (Ours)} & \textbf{77.1} & \textbf{67.8} & \textbf{65.0} \\

            \rowcolor{gainbg}
            
            \textit{Improv. ($\Delta$)} & 
            \textcolor{goodgreen}{$\blacktriangle$ \textbf{+6.3}} & 
            \textcolor{goodgreen}{$\blacktriangle$ \textbf{+8.3}} & 
            \textcolor{goodgreen}{$\blacktriangle$ \textbf{+0.6}} \\
            
            \bottomrule
        \end{tabular}
    }
\end{table}

\begin{figure}[h]
    \centering
    \includegraphics[width=\linewidth]{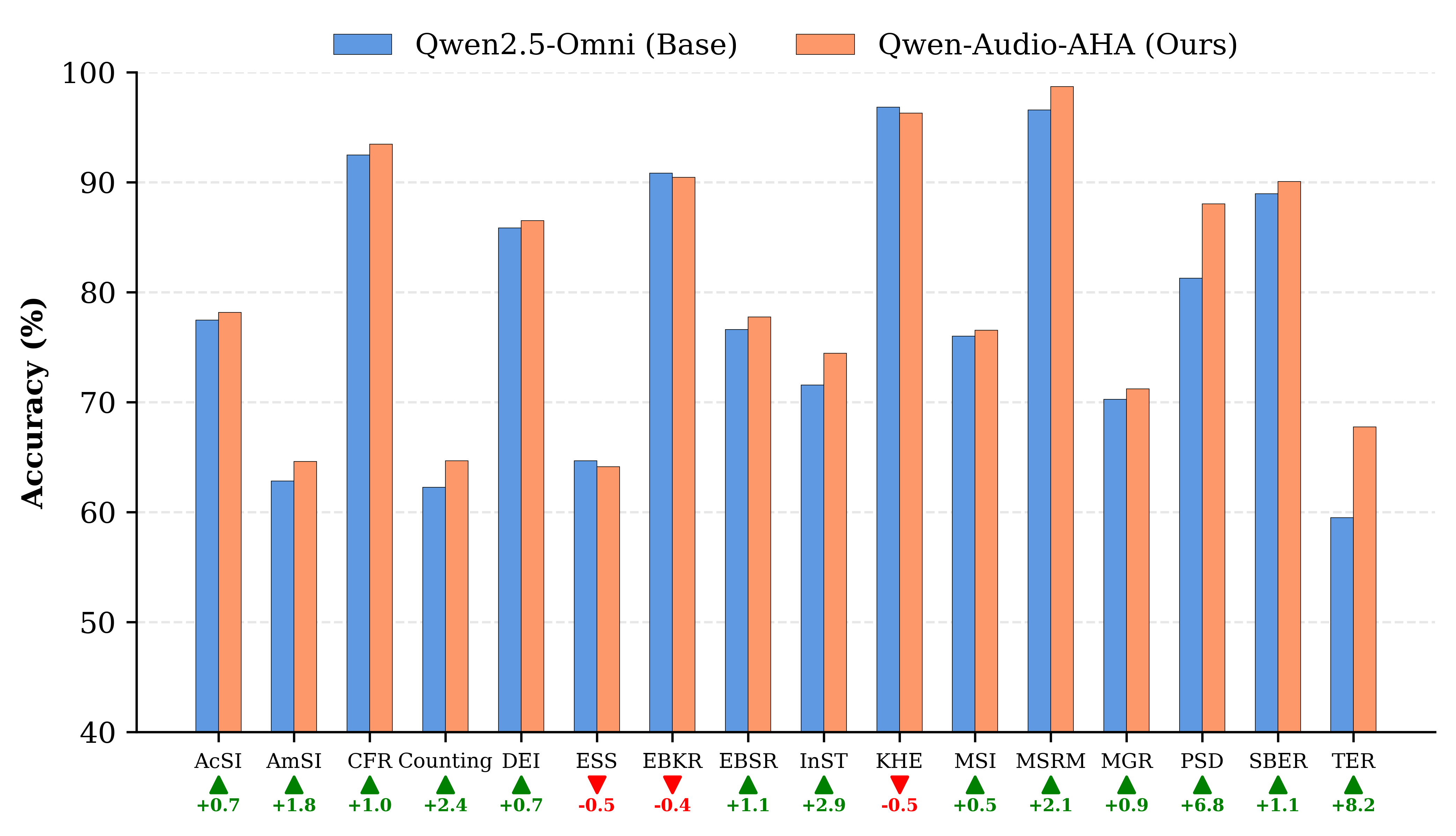}
    \caption{Categorical accuracies before and after alignment in MMAU-test benchmark. Most dimensions have improved accuracy after alignment, especially \textbf{Temporal Event Reasoning (TER)} and \textbf{Phonological Sequence Decoding (PSD)}, while  stagnation or minor degradation are observed in a few subcategories.}
    \label{fig:mmau_dimension}
\end{figure}

\paragraph{\textit{How effective is alignment across different taxonomies?}}
Beyond overall gains, the MMAU-test breakdown (Fig.~\ref{fig:mmau_dimension}) shows improvements across most dimensions, led by \textbf{TER} (+8.2\%) and \textbf{PSD} (+6.8\%).
Subjective categories like \textbf{Emotion State Summarisation (ESS)} (-0.5\%) saw slight drops (note that our alignment data excludes emotion-specific examples).
Crucially, this broad success indicates that the precise temporal grounding enforced by \textbf{AHA} yields a positive transfer effect, enhancing fundamental representation and boosting performance across most taxonomies beyond just specific hallucination mitigation.\textcolor{red}{(For full names of those categories, refer to appendix~\ref{appendix:3})}

% \paragraph{\textit{Does the LLM judge confuse lexical mismatches with genuine errors?}}
% At evaluation time, we leveraged an LLM-as-a-judge method by providing ground-truth captions to a judge model to evaluate the LALM's responses. However, our analysis suggests that the judge model tends toward high strictness and may classify an answer as incorrect even when it is semantically close to the ground truth.

% For instance, if the ground truth specifies "A clicking sound occurred" and the model predicts "There's a ticking sound," the judge model may misclassify this as a False Event Identity error. In such cases, while the model accurately captures the fundamental acoustic properties of the audio wave, the final semantic label often depends on preferences inherited from the pre-training stage, specifically, whether the model chooses to reason one step further to identify the underlying source of the sound event. While our current framework successfully penalizes blatant hallucinations, further investigation into the boundary between lexical mismatching and genuine grounding errors remains a significant direction for future research.

\section{Conclusion}
\label{sec:conclusion}

In this paper, we introduced \textbf{AHA} to mitigate hallucinations in LALMs.
By synthesizing \textbf{counterfactual hard negatives} based on a fine-grained taxonomy, we effectively force models to prioritize acoustic evidence over linguistic priors.
For evaluation, we established \textbf{AHA-Eval}, a diagnostic benchmark tailored for detecting these hallucinations.
Finally, Our aligned model demonstrates significant improvements, even outperforming SOTA methods on both diagnostic and public benchmarks.

\section{Limitations}
\label{sec:limitations}

While \textbf{Qwen-Audio-AHA} demonstrates significant advancements in aligning audio-language models, our study highlights a critical bottleneck in the broader research landscape: the inadequacy of current automated evaluation protocols. 
We observe that even the most advanced models (e.g., GPT-5, Gemini-3) used as judges exhibit rigid sensitivity that often penalizes valid, high-level reasoning.

\begin{figure}[h]
    \centering
    \includegraphics[width=\linewidth]{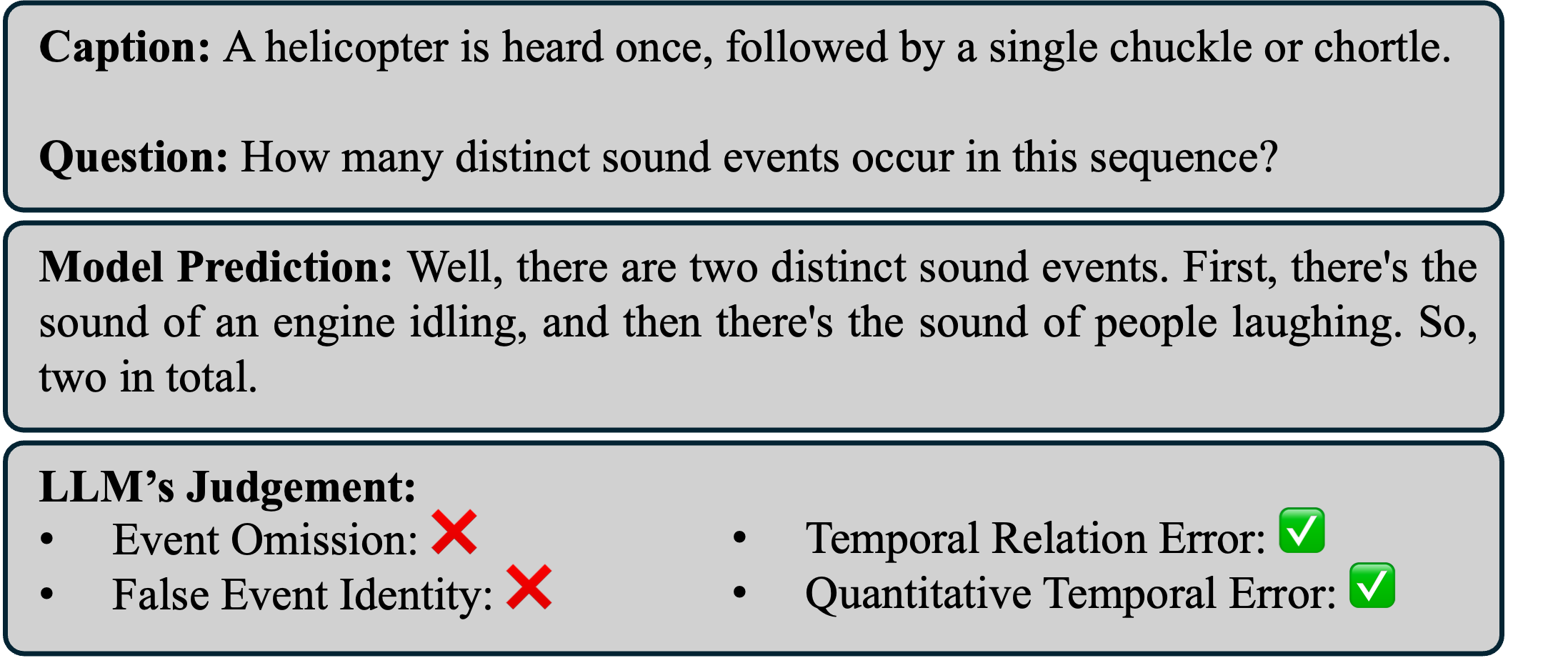}
    \caption{A critical failure case of automated LLM evaluation. The LALM correctly identifies the acoustic event as ``people laughing'', which is semantically equivalent to the caption's ``chuckle or chortle''. However, the LLM judge fails to bridge this semantic gap, incorrectly penalizing the model with \textbf{Event Omission} and \textbf{False Event Identity} errors. This highlights the risk of false positives in current hallucination benchmarks.}
    \label{fig:limitation_reasoning_depth}
\end{figure}

\paragraph{The ``Semantic Equivalence'' Trap in LLM Judging.}
A primary failure mode of current evaluation pipelines is the inability to recognize semantic equivalence across different lexical choices. 
As illustrated in Figure~\ref{fig:limitation_reasoning_depth}, the ground truth describes a sound as a ``chuckle or chortle'', while our model predicts ``people laughing''.
Despite these terms referring to the same fundamental acoustic event, the LLM judge lacks the acoustic intuition to bridge this semantic gap, incorrectly flagging the response as a \textbf{False Event Identity}. 
Similar spurious errors arise with synonyms like ``shout'' versus ``scream'', or ``engine noise'' versus ``idling''. 
This suggests that current reported hallucination rates may be inflated by false positives, where models are penalized not for factual errors, but for describing the truth with different vocabulary than the caption.

\paragraph{Bias Against Reasoning Depth.}
Furthermore, current benchmarks typically favor a specific ``depth'' of description, which usually adheres strictly to the caption's surface forms and restricts the evaluation of multi-dimensional reasoning.
Audio events can be validly described via onomatopoeia (e.g., ``clicking''), the immediate action (``typing''), or the underlying scene (``someone working at a desk''). 
We find that when LALMs provide more granular acoustic details or higher-level scene inferences than the ground truth, they are often unfairly penalized by metrics that rely on rigid n-gram overlap.
This limitation reflects a gap in the community's infrastructure rather than a model deficit.
We argue that the field must move beyond single-dimension accuracy and develop dynamic benchmarks capable of appreciating the diverse but correct interpretations of complex acoustic scenes.

\bibliography{custom}

\appendix

\section{Implementation Details}
\label{appendix:1}
We provide in Table~\ref{tab:hyper_detail} the hyperparameters we used in training Qwen-Audio-AHA:

\begin{table*}[h]
    \centering
    
    \caption{\textbf{Hyperparameters for training Qwen-Audio-AHA}}
    \label{tab:hyper_detail}
    
    \resizebox{\columnwidth}{!}{
        \begin{tabular}{c c}
            \toprule
            
            \textbf{Hyperparameter} & \textbf{value} \\
            \midrule
            
            LoRA rank & 16 \\
            LoRA $\alpha$ & 32 \\
            LoRA dropout & 0.1 \\
            LoRA bias & None \\
            LoRA target layers & r".*model.*(q\_proj|k\_proj|v\_proj|o\_proj|up\_proj|down\_proj|gate\_proj)" \\
            DPO $\beta$ & 0.3 \\
            DPO warmup ratio & 0.03 \\
            n\_epochs & 8 \\
            learning rate & 2e-6 \\
            weight decay & 0 \\
            Adam $\beta1$ & 0.9 \\
            Adam $\beta2$ & 0.95 \\
            dtype & BF16 \\
            per device train batch size & 1 \\
            gradient accumulation steps & 16 \\

            \bottomrule
        \end{tabular}
    }
\end{table*}

\section{AHA Data Generation Details}
\label{appendix:2}
We provide additional details regarding the construction of the AHA training and evaluation datasets. The generation process followed a structured pipeline designed to elicit and penalize specific audio grounding failures.

\subsection{Question Paradigm Selection}
\label{appendix:2.1}
We initially curated a set of candidate question paradigms focusing on event identification, quantitative counting, and complex temporal relationships. These paradigms were synthesized from common query structures found in public open-ended QA datasets, such as \textbf{MECAT-QA} and multiple-choice benchmarks like \textbf{MMAU}. By adapting these existing structures, we ensured that the AHA questions reflect realistic challenges in general audio understanding while maintaining a specialized focus on fine-grained reasoning.

\subsection{Automated Preference Construction}
\label{appendix:2.2}
Using audio-caption pair from the \textbf{AudioTime} corpus as the foundation, we employed \textbf{GPT-4o} to instantiate the shared question pool ($\mathcal{D}$). For each instance, the model was prompted to select an appropriate question template and generate a corresponding preference pair based on the following rules:
\begin{itemize}
    \item \textbf{Chosen Response:} A response that accurately answers the query while remaining strictly grounded in the caption-derived ground truth.
    \item \textbf{Rejected Response:} A response that appears linguistically plausible but deliberately incorporates one or more taxonomy-defined hallucinations: \textbf{Event Omission}, \textbf{False Event Identity}, \textbf{Temporal Relation Error}, or \textbf{Quantitative Temporal Error}.
\end{itemize}

\subsection{Prompting and Quality Control}
\label{appendix:2.3}
\begin{figure*}[h]
    \centering
    \includegraphics[width=\linewidth]{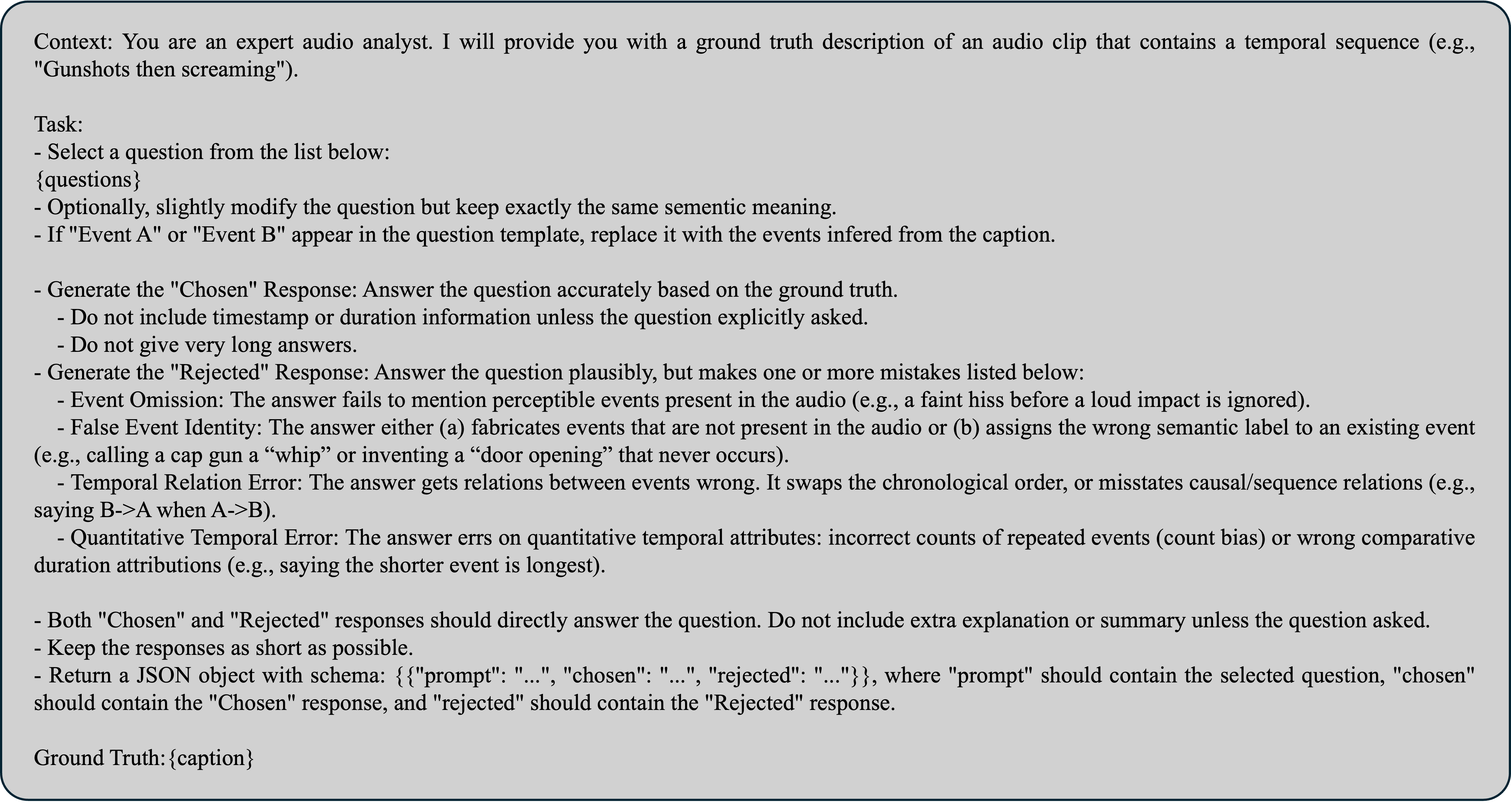}
    \caption{The prompt used for AHA data generation.}
    \label{fig:prompt_data_generation}
\end{figure*}

\begin{figure*}[h]
    \centering
    \includegraphics[width=\linewidth]{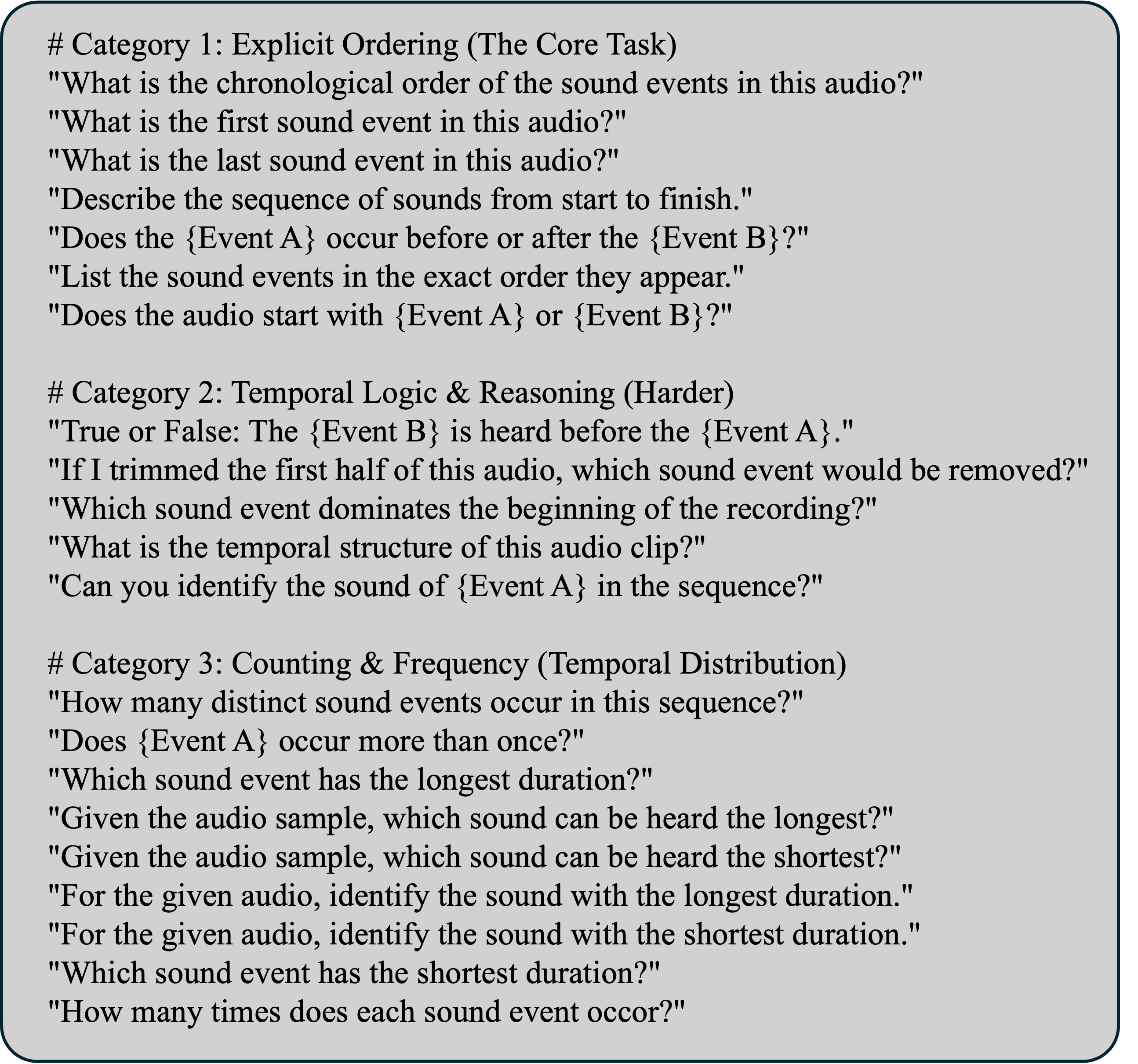}
    \caption{Candidate questions for AHA data generation.}
    \label{fig:candidate_questions}
\end{figure*}

The generation prompt included exhaustive definitions and detailed examples for each hallucination category to guide the judge model toward generating high-fidelity negative samples. The final output was formatted as a structured JSON object to facilitate downstream processing. Following the automated generation phase, human volunteers performed a manual audit of the samples to verify their acoustic accuracy, linguistic plausibility, and categorical diversity. Any instances found to be ambiguous or misaligned with the acoustic evidence were discarded to ensure the integrity of the \textbf{AHA-Eval} benchmark. We present in Figure~\ref{fig:prompt_data_generation} the prompt and Figure~\ref{fig:candidate_questions} candidate questions we used for AHA data generation.

\section{MMAU-test benchmark Category Abbreviations}
\label{appendix:3}
For aesthetics reason, we put abbreviations in our categorical demonstration of MMAU-test result. We list the names of those abbreviations proposed by the official MMAU-test benchmark in Table~\ref{tab:mmau_category}.

\begin{table*}[htbp]
    \centering
    
    \caption{\textbf{MMAU-test benchmark Category Abbreviations.}}
    \label{tab:mmau_category}
    
    \resizebox{\columnwidth}{!}{
        \begin{tabular}{c c}
            \toprule
            
            \textbf{Abbreviation} & \textbf{Name} \\
            \midrule
            
            AcSI & Acoustic Source Inference \\
            AmSI & Ambient Sound Interpretation \\
            CFR & Conversational Fact Retrieval \\
            Counting & Counting \\
            DEI & Dissonant Emotion Interpretation \\
            ESS & Emotion State Summarisation \\
            EBKR & Event-Based Knowledge Retrieval \\
            EBSR & Event-Based Sound Reasoning \\
            InST & Instrumentation \\
            KHE & Key highlight Extraction \\
            MSI & Melodic Structure Interpretation \\
            MSRM & Multi Speaker Role Mapping \\
            MGR & Musical Genre Reasoning \\
            PSD & Phonological Sequence Decoding \\
            SBER & Sound-Based Event Recognition \\
            TER & Temporal Event Reasoning \\
            
            \bottomrule
        \end{tabular}
    }
\end{table*}
% \label{appendix:2}

\end{document}